\documentclass[10pt,conference]{IEEEtran}

\usepackage{cite}
\usepackage{subfigure}
\usepackage{graphicx}
\usepackage{url}
\usepackage{amsmath}
\usepackage{amssymb}
\begin{document}

\title{On packet lengths and overhead for random linear coding over the erasure channel}

\author{
\authorblockN{Brooke Shrader and Anthony Ephremides}
\authorblockA{Electrical and Computer Engineering Dept\\
and Institute for Systems Research \\
University of Maryland \\
College Park, MD 20742 \\
bshrader, etony@umd.edu} }

\maketitle

\begin{abstract}
We assess the practicality of random network coding by illuminating
the issue of overhead and considering it in conjunction with
increasingly long packets sent over the erasure channel. We show
that the transmission of increasingly long packets, consisting of
either of an increasing number of symbols per packet or an
increasing symbol alphabet size, results in a data rate approaching
zero over the erasure channel. This result is due to an erasure
probability that increases with packet length. Numerical results for
a particular modulation scheme demonstrate a data rate of
approximately zero for a large, but finite-length packet. Our
results suggest a reduction in the performance gains offered by
random network coding.
\end{abstract}

\section{Introduction}

It has been shown that network coding allows a source node to
multicast information at a rate approaching the maximum flow of a
network as the symbol alphabet size approaches infinity
\cite{AhlswedeEtal00, LiYeungCai03}. An infinitely large symbol
alphabet would correspond to transmission of infinitely long blocks
of data, which is neither possible nor practical. Subsequent works
have shown that a finite alphabet is sufficient to achieve the
maximum flow and a number of works have provided bounds on the
necessary alphabet size. For example, in \cite{KoetterMedard02} the
alphabet size is upper bounded by the product of the number of
sources with the number of receivers, while in \cite{JaggiEtal05} a
lower bound on the alphabet size is given by the number of
receivers.

Random linear coding was proposed to allow for distributed
implementation \cite{HoEtal03}. In random network coding, a randomly
generated encoding vector is typically communicated to the receiver
by appending it to the header of the transmitted packet. The
overhead inherent in communicating the encoding vector becomes
negligible as the number of symbols in the packet grows large.
Regarding the alphabet size for random network coding, a lower bound
of 2 times the number of receivers is given in \cite{JaggiEtal05}.

The ``length'' of a transmitted packet (i.e., the number of bits
conveyed by the packet) depends on both the symbol alphabet and the
number of symbols per packet. From the previous works listed above,
it is clear that transmitting sufficiently long packets is crucial
to ensuring the existence of network codes and to allowing random
linear coding to operate with low overhead.

However, long packets are more susceptible to noise, interference,
congestion, and other adverse channel effects. This point has been
ignored in previous works on network coding. In particular, a number
of previous works \cite{LunEtal, EryilmazEtal06, DanaEtal06}
consider transmission over an erasure channel, in which a packet is
either dropped with a probability $\epsilon$ or received without
error. In a wireline network, the erasure channel is used to model
dropped packets due to buffer overflows. Clearly, longer packets
take up more space in memory, so the erasure probability will
increase as the packet length grows. In the case of a wireless
network, there are a number of reasons that the erasure probability
will increase with the length of the packet, as listed below.
\begin{itemize}
\item If we try to fit more bits into the channel using modulation,
then for fixed transmission power, points in the signal
constellation will move closer together and errors are more likely.
\item If we try to fit more bits into the channel by decreasing
symbol duration, then we are constrained by bandwidth, a
carefully-controlled resource in wireless systems.
\item Longer packets are more susceptible to the effects of fading.
\end{itemize}

In this work, we model the erasure probability as a function of the
packet length and investigate the implications on network coding
performance. This is somewhat reminiscent of
\cite{SmithVishwanath06}, in which the erasure probability is a
function of the link distance in a wireless network, and the effect
on capacity is investigated. Our emphasis will be on a wireless
channel with a fixed bandwidth, for which we will associate erasure
probability with the probability of symbol error for a given
modulation scheme.

We note that a careful examination of packet lengths in data
networks is not a new idea; most notably, many researchers
participated in a dispute over the packet size for the Asynchronous
Transfer Mode (ATM) standard in the 1980s. And still many tradeoffs
are being studied today at the network level regarding the packet
length. The interplay between packet lengths and coding arises
because of the way that network coding unifies different layers of
the protocol stack. Another work which examines a similar problem is
\cite{HongNosratinia02}, in which packet headers and packet lengths
are analyzed for Reed-Solomon coding in video applications.

\section{Throughput of random linear coding}

We consider the following setting. A source node has $K$ units of
information $\{s_1,s_2,\hdots,s_K\}$ that it wants to transmit. We
will refer to each of these information units as packets and let
each packet be given by a vector of $n$ $q$-ary symbols, where $q$
is the symbol alphabet (the size of a finite field) and $n$ is the
number of symbols per packet. We consider values of $q$ which are
powers of 2, i.e., $q=2^u$ for some $u \geq 1$. The length of a
packet is given by $n \log_2 q$ bits. In Section
\ref{section:noprecode} we will consider the case where
$\{s_1,s_2,\hdots,s_K\}$ are packets of uncoded information, whereas
in Section \ref{section:precode} we will assume that a
(deterministic) error-correcting code has been applied in order to
form $\{s_1,s_2,\hdots,s_K\}$.

The source generates random linear combinations by forming the
modulo-$q$ sum $\sum_{i=1}^{K} \alpha_i s_i$, where $\alpha_i$ are
chosen randomly and uniformly from the set $\{0,1,2,\hdots,q-1\}$.
Note that the resulting random linear combination is a packet with
length $n \log_2 q$ bits. With each random linear combination
transmitted, the source appends a packet header identifying
$\alpha_i$, which requires an additional $K \log_2 q$ bits of
overhead with every $n \log_2 q$ bits transmitted. A receiver will
collect these random linear combinations until it has received
enough to decode. For transmission over an erasure channel, decoding
can be performed by Gaussian elimination. Let $N$ denote the number
of random linear combinations needed for decoding. Each of the $N$
random linear combinations represents a linear equation in
$\{s_1,s_2,\hdots,s_K\}$ and the distribution of $N$ is given by
\begin{equation*}
\mbox{Pr}(N{\leq}j) \!=\! \mbox{Pr}\{ \mbox{a random $K {\times} j$
matrix has rank } K\}.
\end{equation*}
Note that Pr$(N \leq j)$ is equal to zero for $j<K$. For $j \geq K$,
the distribution can be found following the procedure in
\cite{MacKay98} and is given by
\begin{eqnarray*}
\mbox{Pr}(N{\leq}j) & \!\!=\!\! &
\frac{(q^j{-}1)(q^j{-}q)(q^j{-}q^2)\hdots(q^j{-}q^{K{-}1})}{q^{jK}} \\
 & \!\!=\!\! & \prod_{i=0}^{K-1} (1-q^{-j{+}i})
\end{eqnarray*}
The expected value of $N$ can be found from the above distribution
and is shown to be given by
\begin{eqnarray}
E[N] & = & K + \sum_{j=K}^{\infty} \left( 1 -
\prod_{i=0}^{K-1}(1-q^{-j{+}i})  \right) \nonumber \\
 & = & \sum_{i=1}^{K}(1-q^{-i})^{-1} \label{eqn:EN}
\end{eqnarray}
Clearly, as $q \rightarrow \infty$, $E[N] \rightarrow K$.

We let $\epsilon(n \log_2 q)$ denote the erasure probability on the
channel, which is an increasing function of the packet length $n
\log_2 q$. The expected number of transmissions needed for the
receiver to decode the original $K$ packets is given by
\begin{equation*}
\frac{E[N]}{1-\epsilon(n \log_2 q)}.
\end{equation*}
Over the course of these transmissions, the average number of
packets received for each transmission is
\begin{equation*}
\frac{K(1-\epsilon(n \log_2 q))}{E[N]}
\end{equation*}
We account for the overhead by scaling the number of packets
received per transmission by $n/(n + K)$, which is the ratio of
number of information symbols to the total number of symbols
(information plus overhead) sent with each transmission. We define
$S$ as the effective portion of each transmission which contains
message information; $S$ is a measure of throughput in packets per
transmission and takes values between 0 and 1.
\begin{equation}
S = \frac{K}{E[N]} \frac{n}{n+K} (1-\epsilon(n \log_2 q))
\label{eqn:S}
\end{equation}
If the erasure probability is constant, $\epsilon(n \log_2 q) = e$,
then for $n$, $q$ $\rightarrow \infty$, corresponding to infinitely
long packets, the value of $S$ approaches $1-e$, which is the
Shannon capacity of the erasure channel.

From the expression in (\ref{eqn:S}), we can identify a tradeoff
between the packet length and the throughput. As the number of
information symbols per packet $n$ grows large, the effect of
overhead becomes negligible, but the erasure probability grows and
the transmissions are more likely to fail. Alternatively, if $n$
approaches zero, transmissions are more likely to succeed, but the
overwhelming amount of overhead means that no information can be
transmitted. In a similar manner, as the alphabet size $q$ grows,
the random linear coding becomes more efficient in the sense that
$E[N] \rightarrow K$, but again, the erasure probability increases
and transmissions are likely to be unsuccessful. This tradeoff
demonstrates that the alphabet size, packet length, and overhead
must be carefully weighed in determining the performance of random
linear coding over the erasure channel.

\section{Performance without pre-coding} \label{section:noprecode}

In this section we consider the case where $\{s_1, s_2,\hdots,s_K\}$
are uncoded information symbols. We define the data rate in bits per
transmission as $R$, where
\begin{equation}
R = S n \log_2 q. \label{eqn:Rnoprecode}
\end{equation}
The data rate $R$ accounts for the fact that each received packet
contains $n \log_2 q$ bits of information. Since the packets $\{s_1,
s_2,\hdots,s_K\}$ consist of uncoded information, for every random
linear combination sent, all $n$ symbols must be received without
error. Then
\begin{equation*}
\mbox{Pr}(\mbox{erasure}) = 1-\left( 1- \mbox{Pr}(\mbox{symbol
error})\right)^{n}
\end{equation*}
and $\mbox{Pr}(\mbox{symbol error})$ will correspond to the symbol
error probability for $q$-ary modulation over the channel. We denote
the probability of symbol error by $P_q$ and note that it is
independent of $n$ but depends on $q$ as well as features of the
channel such as pathloss, signal-to-noise ratio (SNR), and fading
effects. We will consider a wireless channel of limited bandwidth,
for which modulation techniques such as pulse amplitude modulation
(PAM), phase shift keying (PSK), and quadrature amplitude modulation
(QAM) are appropriate. For these modulation techniques, $P_q
\rightarrow 1$ as $q \rightarrow \infty$ \cite{Proakis}.

In this setting, the erasure probability is given by
\begin{equation}
1-\epsilon(n \log_2 q) = (1-P_q)^{n}.
\end{equation}
From the above expression we note that, as suggested by our
discussion on the tradeoff between throughput and packet length, as
$n \rightarrow \infty$, both $S$ and $R$ approach zero exponentially
fast. Additionally, for the modulation schemes mentioned above, $S$
and $R$ approach zero as $q \rightarrow \infty$, albeit at a slower
rate.

\begin{figure}
\centering
\includegraphics[width=3.2in]{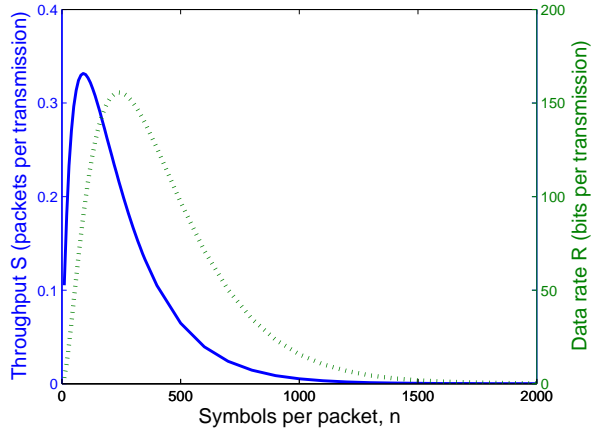}
\caption{Throughput $S$ (solid line) and data rate $R$ (dotted line)
versus symbols per packet $n$ without pre-coding for QAM modulation,
$K=80$, $q=8$, and SNR per bit $\gamma_b$=3.5dB.}
\label{fig:SandRVsLengthNoPC}
\end{figure}

\begin{figure}
\centering
\includegraphics[width=3.2in]{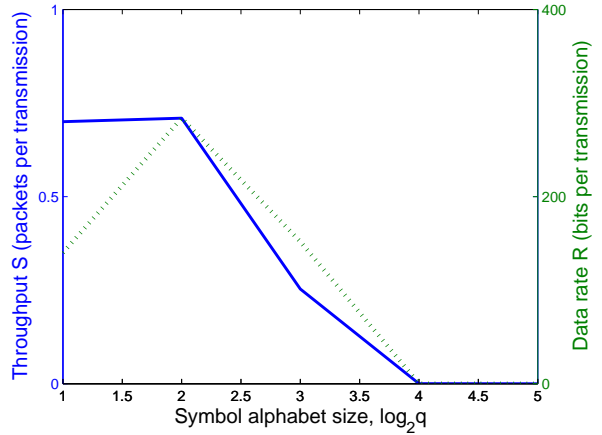}
\caption{Throughput $S$ (solid line) and data rate $R$ (dotted line)
versus log of the symbol alphabet $\log_2q$ for QAM modulation,
$K=80$, $n=200$, SNR per bit $\gamma_b$=3.5dB, and no pre-coding.}
\label{fig:SandRVsAlphabetNoPC}
\end{figure}

As a numerical example, we have plotted the throughput $S$ and data
rate $R$ for QAM modulation over an additive white Gaussian noise
(AWGN) channel. In this case the symbol error probability for the
optimum detector is approximated by \cite{Proakis}
\begin{equation}
P_q \approx 1 - \left( 1 - 2\left(1-\frac{1}{\sqrt{q}} \right)
Q\left( \frac{3 \gamma_b \log_2 q}{q-1}\right)\right)^2
\label{eqn:PqQAM}
\end{equation}
where $\gamma_b$ is the SNR per bit and $Q$ is the complementary
cumulative distribution function for the Gaussian distribution. The
above expression for $P_q$ holds with equality for $\log_2 q$ even.
The results are shown in Figure \ref{fig:SandRVsLengthNoPC}, where
$q$ is fixed and $n$ is varied, and in Figure
\ref{fig:SandRVsAlphabetNoPC}, where $n$ is fixed and $q$ is varied.
In all cases, the throughput and data rate are concave functions,
admitting optimum values for $n$ and $q$. Furthermore, if $n$ grows
large as $q$ is fixed or if $q$ grows large as $n$ is fixed, the
throughput approaches zero.

To improve the performance described above, a number of techniques
might be employed, including enhanced receivers, diversity
techniques, and channel coding.

%\begin{figure*}
%\centerline{\subfigure[Erasure probability increasing
%exponentially.]{\includegraphics[width=3.5in]{figs/SVsQExp.eps}
%\label{fig:SVsQExp}} \hfil \subfigure[Erasure probability increasing
%polynomially.]{\includegraphics[width=3.5in]{figs/SVsQPoly.eps}
%\label{fig:SVsQPoly}}} \caption{Throughput $S$ versus alphabet size
%$q$ for fixed $K$ and $\lambda$.} \label{fig:SVsQ}
%\end{figure*}

\section{Performance with pre-coding} \label{section:precode}

In this section we examine the effect of using a deterministic
pre-code prior to performing random linear coding. We assume that an
uncoded packet $\sigma_i$, consisting of $k$ $q$-ary information
symbols, is passed through an encoder to add redundancy, producing
the packet $s_i$ consisting of $n$ $q$-ary symbols. Thus a $q$-ary
code of rate $R_{pc}=k/n$ is used to produce
$\{s_1,s_2,\hdots,s_K\}$. Random linear coding is again performed on
$\{s_1,s_2,\hdots,s_K\}$ and we assume that the codebook for the
pre-code is known to the decoder and does not need to be transmitted
over the channel. At the receiver, decoding of the random linear
code will be performed first using the coefficients $\alpha_i$,
$i=1,2,\hdots, K$ sent with each random linear combination.
Subsequently the pre-code will be decoded. The use of the pre-code
means that the system will be tolerant to some errors in the $n$
symbols sent in each random linear combination.

To account for the pre-code, we modify the throughput $S$ expressed
in (\ref{eqn:S}) by multiplying it be $R_{pc}$, which results in
\begin{equation}
S = \frac{K}{E[N]} \frac{k}{n+K} (1-\epsilon(n \log_2 q)).
\label{eqn:Sprecode}
\end{equation}
To obtain the data rate $R$ in bits per transmission, $S$ is now
scaled by $k$, the number of information symbols present in each
packet.
\begin{equation}
R = S k \log_2 q
\end{equation}

We can bound the erasure probability, and thus the throughput $S$
and data rate $R$, using bounds on the error-correcting capabilities
of the pre-code. Let $t$ denote the number of symbol errors that the
pre-code can correct. For a given code, $t = \lfloor
 1/2(d_{min}-1)\rfloor$ errors, where $d_{min}$ is the minimum distance of the code.
The erasure probability can be bounded as follows.
\begin{eqnarray*}
1 - \epsilon(n \log_2 q) & \stackrel{(a)} \geq & 1 -
\sum_{i=t+1}^{n}
\mbox{Pr}(i\mbox{ symbols in error}) \\
 & \stackrel{(b)} = & 1 - \sum_{i=t+1}^{n} \binom{n}{i} P_q^i (1-P_q)^{n -
 i} \\
& \stackrel{(c)} \geq & \sum_{i=0}^{ \lfloor
 1/2(d-1)\rfloor } \binom{n}{i} P_q^i (1-P_q)^{n -
 i}
\end{eqnarray*}
In the above expression, $(a)$ holds with equality for the class of
perfect codes \cite{Proakis}, $(b)$ follows by assuming that errors
are independent, identically distributed $\sim P_q$ among the $n$
symbols, and $(c)$ holds for $d \leq d_{min}$. We make use of the
Gilbert-Varshamov bound to provide a lower bound on $d_{min}$ as
follows.
\begin{equation}
d = \inf \left\{d_{min}: \sum_{i=0}^{d_{min}-2}
\binom{n-1}{i}(q-1)^{i} \geq q^{n-k} \right\}
\end{equation}

Lower bounds on the throughput and data rate are given as follows.
\begin{equation}
S_{LB} = \frac{K}{E[N]} \frac{k}{n+K} \sum_{i=0}^{ \lfloor
 1/2(d-1)\rfloor } \binom{n}{i} P_q^i (1-P_q)^{n -
 i}
\end{equation}
\begin{equation}
R_{LB} = S_{LB} k \log_2 q
\end{equation}
For $q \rightarrow \infty$ while all other parameters are fixed, if
the modulation scheme has $P_q \rightarrow 1$, then $S_{LB}, R_{LB}
\rightarrow 0$. In the limit of increasing symbols per packet $n$,
we identify two different cases. First, if the pre-code rate
$R_{pc}$ is fixed (i.e, $k \rightarrow \infty$ as $n \rightarrow
\infty$ with fixed ratio between the two) then $S_{LB} \rightarrow
R_{pc}$ (multiplied by $K/E[N] \approx 1)$ and $R_{LB}$ increases
without bound. On the other hand, if $k$ is fixed (i.e., as $n
\rightarrow \infty$, $R_{pc} \rightarrow 0$) then $S_{LB}, R_{LB}
\rightarrow 0$.

\begin{figure}
\centering
\includegraphics[width=3.2in]{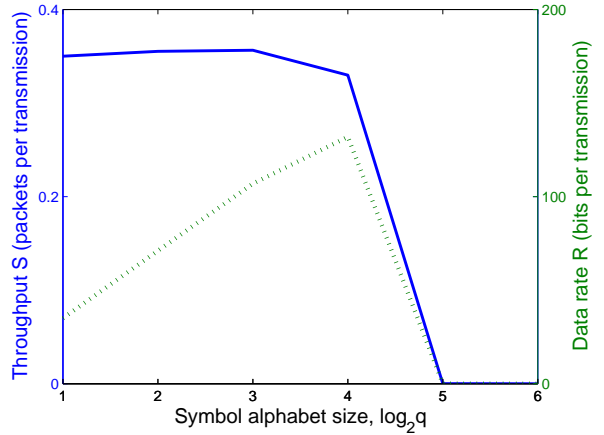}
\caption{Bounds on the throughput $S_{LB}$ (solid line) and data
rate $R_{LB}$ (dotted line) versus log of the symbol alphabet
$\log_2q$ for QAM modulation, $K=80$, $n=200$, $k=100$ (pre-code
rate 1/2), and SNR per bit $\gamma_b$=3.5dB.}
\label{fig:SandRVsAlphabetWithPC}
\end{figure}

We have computed numerical examples of $S_{LB}$ and $R_{LB}$
assuming QAM modulation with $P_q$ as given in (\ref{eqn:PqQAM}). In
Fig. \ref{fig:SandRVsAlphabetWithPC} we display the throughput and
data rate as a function of $q$. Clearly, pre-coding allows the
system to support higher $q$ with non-zero throughput, although the
throughput does eventually go to zero for large $q$. In comparing
this result to Fig. \ref{fig:SandRVsAlphabetNoPC}, note that all
other parameters are fixed and the pre-code rate $R_{pc}= 1/2$
results in a halving of the throughput and data rate.

\begin{figure*}
\centerline{\subfigure[Fixed rate pre-code
$R_{pc}=1/2$.]{\includegraphics[width=3.2in]{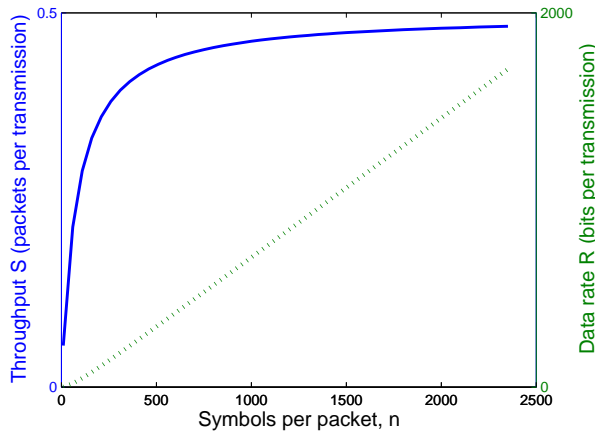}
\label{fig:SandRVsLengthWithPCFixedRate}} \hfil \subfigure[Variable
pre-code rate,
$k=400$.]{\includegraphics[width=3.2in]{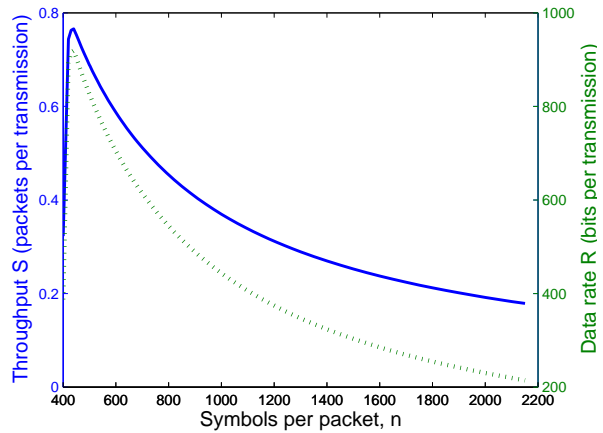}
\label{fig:SandRVsLengthWithPCVaryRate}}} \caption{Bounds on the
throughput $S_{LB}$ (solid line) and data rate $R_{LB}$ (dotted
line) versus symbols per packet $n$ for QAM modulation, $K=80$,
$q=8$, and SNR per bit $\gamma_b$=3.5dB.}
\label{fig:SandRVsLengthWithPC}
\end{figure*}

In Fig. \ref{fig:SandRVsLengthWithPC} we have displayed the two
different cases for performance as a function of increasing symbols
per packet $n$. In Fig. \ref{fig:SandRVsLengthWithPCFixedRate}, the
rate $R_{pc}$ is fixed at $1/2$ and $S_{LB} \rightarrow R_{pc}$
while $R_{LB}$ increases without bound. In this case, pre-coding
ensures that the throughput and data rate increase with $n$.
However, there is a caveat: the result in Fig.
\ref{fig:SandRVsLengthWithPCFixedRate} requires that $k \rightarrow
\infty$, which means that the source must have an infinite supply of
uncoded information to be transmitted. On the other hand, if the
source has a finite amount of information to be transmitted, as
shown in Fig. \ref{fig:SandRVsLengthWithPCVaryRate}, then as $n
\rightarrow \infty$, the pre-code rate $R_{pc} \rightarrow 0$ and
the data rate also approaches zero. In summary, the use of a
pre-code can improve performance on the erasure channel; indeed,
pre-coding is the mechanism which allows us to represent a noisy
channel as an erasure channel. However, pre-coding can only ensure
that the data rate increases with packet length if there is an
infinite supply of (uncoded) information at the source. This is a
strong assumption for a network, where data often arrives in bursts.

%%%==========tried to balance pages, but didn't make this work=============
\IEEEtriggercmd{\enlargethispage{-5in}}

\section{Discussion}

The primary contribution of this work is the elucidation of the fact
that the widely-used assumption of the erasure probability being
constant as the packet length increases leads to idealized
performance that cannot be obtained in a practical scenario. We have
shown that the assumption of increasingly long packets, either due
to increasingly many symbols per packet or to an increasing alphabet
size, can result in a data rate of zero for random linear coding if
the erasure probability increases with packet length. While the use
of a pre-code prior to performing random linear coding can improve
performance, it can only ensure that the throughput increases with
packet length if there is an infinite supply of data awaiting
transmission at the source. We have also demonstrated that the
overhead needed for random linear coding can have a devastating
impact on the throughput. We have focused our attention on the
erasure channel, but the tradeoff between packet length and
throughput will arise in other channels, particularly in wireless
channels where these effects can be exacerbated by fading. A number
of techniques, such as pre-coding, can be employed to combat the
adverse effects of the channel, but in making use of these
techniques, the performance gains offered by network coding will be
tempered.

\section*{Acknowledgment}
% optional entry into table of contents (if used)
%\addcontentsline{toc}{section}{Acknowledgment}
This work is supported by the Office of Naval Research through grant
N000140610065, by the Department of Defense under MURI grant
S0176941, and by NSF grant CNS0626620. Prepared through
collaborative participation in the Communications and Networks
Consortium sponsored by the U.S. Army Research Laboratory under The
Collaborative Technology Alliance Program, Cooperative Agreement
DAAD19-01-2-0011. The U.S. Government is authorized to reproduce and
distribute reprints for Government purposes notwithstanding any
copyright notation thereon. The views and conclusions contained in
this document are those of the authors and should not be interpreted
as representing the official policies, either expressed or implied,
of the Army Research Laboratory or the U. S. Government.

%\IEEEtriggeratref{11}


% http://www.ctan.org/tex-archive/macros/latex/contrib/supported/IEEEtran/bibtex
%\bibliographystyle{IEEEtran.bst}

\end{document}